\documentclass[aps,preprint,nofootinbib,floatfix]{revtex4-1}
\usepackage{xcolor}
\usepackage{graphicx}
\usepackage[latin1]{inputenc}
\usepackage{amsmath,amssymb}
\usepackage{hyperref}
\usepackage{epstopdf}

\newcommand{\lsim}{\mathrel{\mathop{\kern 0pt \rlap
  {\raise.2ex\hbox{$<$}}}
  \lower.9ex\hbox{\kern-.190em $\sim$}}}
\newcommand{\gsim}{\mathrel{\mathop{\kern 0pt \rlap
  {\raise.2ex\hbox{$>$}}}
  \lower.9ex\hbox{\kern-.190em $\sim$}}}

\newcommand{\gev}{\ensuremath{\,\mathrm{GeV}}}

\def  \bcen   {\begin{center}}
\def  \ecen   {\end{center}}
\def  \beq    {\begin{equation}}
\def  \eeq    {\end{equation}}
\def  \bpm    {\begin{pmatrix}}
\def  \epm    {\end{pmatrix}}
\def  \beqa   {\begin{eqnarray}}
\def  \eeqa   {\end{eqnarray}}
\def  \nn     {\nonumber }
\def\bea{\begin{eqnarray}}
\def\eea{\end{eqnarray}}

\def\ga   {\gamma}
\def\Ga   {\Gamma}
\def\th   {\theta}
\def\la   {\lambda}

\def\sig   {\sigma}

\def\nn{\nonumber}
\def\lee { \left( }
\def\rii { \right) }
\def\lan   {\langle}
\def\ran   {\rangle}
\def\de {\delta}
\def\De {\Delta}

\def\to {\rightarrow}
\def\mlh {m_ {\textsc{\tiny{$\ell^H$}} }}

\begin{document}

{\small
\begin{flushright}
IPMU15-0219  \; DO-TH 15/18 \\
\end{flushright} }

\title{Gauged Two Higgs Doublet Model\\ 
Confronts The LHC 750 GeV Diphoton Anomaly }
\author{
Wei-Chih Huang$^{1}$, Yue-Lin Sming Tsai$^2$, Tzu-Chiang Yuan$^{3,4}$
}

\affiliation{
\small{
$^1$Fakult\"at f\"ur Physik, Technische Universit\"at Dortmund,
44221 Dortmund, Germany \\
$^2$Kavli IPMU (WPI), University of Tokyo, Kashiwa, Chiba 277-8583, Japan\\
$^3$Institute of Physics, Academia Sinica, Nangang, Taipei 11529, Taiwan\\
$^4$Physics Division, National Center for Theoretical Sciences, Hsinchu, Taiwan
}
}

\date{\today}

\begin{abstract}
In light of the recent 750 GeV diphoton anomaly observed at the LHC, we study the possibility of accommodating
the deviation from the standard model prediction based on the recently proposed Gauged Two Higgs Doublet Model.
The model embeds two Higgs doublets 
into a doublet of a non-abelian gauge group $SU(2)_H$, while
the standard model $SU(2)_L$ right-handed fermion singlets are paired up with new heavy fermions to form $SU(2)_H$ doublets, 
and $SU(2)_L$ left-handed fermion doublets are singlets under $SU(2)_H$.
An $SU(2)_H$ scalar doublet, which provides masses to the new heavy fermions 
as well as the $SU(2)_H$ gauge bosons, can be produced
via gluon fusion and subsequently decays into two photons with the new fermions
circulating the triangle loops to account for the deviation from the standard model prediction. 

\end{abstract}
%\pacs{}

\maketitle

%%#######################################################%%
\section{Introduction \label{section:1}}
%%#######################################################%%

Recent results from LHC~\cite{ATLASga,CMS:2015dxe,ATLASga1} exhibit an intriguing anomaly on the diphoton channel at the scale around 750 GeV.
Numerous attempts~\cite{Mambrini:2015wyu,
Buttazzo:2015txu,McDermott:2015sck,Cao:2015pto,Fichet:2015vvy,Alves:2015jgx,Harigaya:2015ezk,Molinaro:2015cwg,
Falkowski:2015swt,Backovic:2015fnp,Angelescu:2015uiz,Franceschini:2015kwy,Dutta:2015wqh,deBlas:2015hlv,
Kobakhidze:2015ldh,Chao:2015ttq,Curtin:2015jcv,Benbrik:2015fyz,Cox:2015ckc,Higaki:2015jag,
Pilaftsis:2015ycr,Megias:2015ory,Petersson:2015mkr,Demidov:2015zqn,Bellazzini:2015nxw,Low:2015qep,
No:2015bsn,Bian:2015kjt,Bai:2015nbs,Nakai:2015ptz,DiChiara:2015vdm,Chakrabortty:2015hff,Gabrielli:2015dhk,
Csaki:2015vek,Kim:2015ron,Han:2015cty,Cao:2015twy,Knapen:2015dap,Bernon:2015abk,Becirevic:2015fmu,
Carpenter:2015ucu,Aloni:2015mxa,Ahmed:2015uqt,Agrawal:2015dbf,Boucenna:2015pav,Murphy:2015kag,
Chao:2015nsm,Arun:2015ubr,Basso:2015aee,Chang:2015bzc,Chakraborty:2015jvs,
Ding:2015rxx,Han:2015dlp,Han:2015qqj,Luo:2015yio,Chang:2015sdy,Bardhan:2015hcr,Basso:2012nh,
Feng:2015wil,Antipin:2015kgh,Wang:2015kuj,Huang:2015evq,Liao:2015tow,Heckman:2015kqk,
Dhuria:2015ufo,Bi:2015uqd,Kim:2015ksf,Berthier:2015vbb,Cho:2015nxy,Cline:2015msi,Gupta:2015zzs,
Bauer:2015boy,Chala:2015cev,Barducci:2015gtd,Ellis:2015oso,Hernandez:2015ywg,Dey:2015bur,Dev:2015isx}
have been put forward to explain the excess,
while Refs.~\cite{Angelescu:2015uiz,Han:2015qqj,Becirevic:2015fmu} are based on two Higgs doublet models, similar to this work.

In Ref.~\cite{Ellis:2015oso}, a combined result from run I and II 
gives a cross section $\sigma(pp\to X\to\ga\ga)\sim \mathcal{O}(6)$ fb 
for a  scalar particle $X$  with mass around 750\gev. 
In this paper, we will show that the newly proposed
Gauged Two Higgs Doublet Model~\cite{Huang:2015wts}~(G2HDM) 
is able to provide a cross section with such magnitude.

G2HDM contains additional $SU(2)_H \times U(1)_X$ gauge symmetry, in which $H_1$ 
(identified as the Standard Model (SM) Higgs doublet) and $H_2$ comprise an $SU(2)_H$ doublet such that the two-doublet potential is as simple as the SM Higgs potential with just a quadratic mass term plus a quartic term. 
The cost to pay is to include additional scalars: one $SU(2)_H$ triplet $\De_H$ and one $SU(2)_H$ doublet $\Phi_H$ (that are all singlets under the SM gauge groups) with their vacuum expectation values~(vevs) supplying masses 
to the $SU(2)_H \times U(1)_X$ gauge bosons. 
Moreover, the vev of the triplet induces the SM Higgs vev, breaking $SU(2)_L\times U(1)_Y$ down to $U(1)_Q$, while $H_2$
does not obtain any vev and the neutral component of $H_2$ could be a dark matter (DM) candidate, 
whose stability is protected by the $SU(2)_H$ gauge symmetry and Lorentz invariance,
without resorting to an ad-hoc $Z_2$ symmetry.
In order to write down $SU(2)_H\times U(1)_X$ invariant Yukawa couplings, we introduce heavy $SU(2)_L$ singlet Dirac fermions, the right-handed component of which is paired up with the SM right-handed fermions to comprise $SU(2)_H$ doublets.
In this setup, the model is anomaly-free regarding all gauge groups involved.

In this work, we focus on the role of $\phi_2$ which is a physical component in $\Phi_H$ and whose vev $\lan \phi_2 \ran = v_\Phi$ gives masses to the new heavy fermions.
Since it couples to new colored fermions, it can be produced radiatively via gluon fusion and also decay radiatively into a pair of photons with the heavy charged fermions in loops.
We will demonstrate that $\phi_2$ can be a good candidate 
if LHC eventually could confirm the diphoton anomaly.  
Moreover, the observed width of the bump can be simply obtained 
from $\phi_2$ decay into the additional fermions
with $\mathcal{O}(1)$ Yukawa couplings.

The paper is organized as follows. First, we briefly discuss the G2HDM in Section~\ref{section:model} 
restraining ourselves only to those aspects most relevant to $\ga\ga$ mode.
Next, in Section~\ref{section:diph} we compute the diphoton cross section through $\phi_2$ exchange and the partial decay width of $\phi_2$ into the new heavy fermions. 
In Section~\ref{section:Hfermion}, we briefly comment on 
implications of such the new heavy fermions in terms of collider searches, 
electron and muon magnetic dipole moment measurements, 
and the electroweak precision test
data. Finally, we conclude in Section~\ref{section:conclusion}.       

%%#######################################################%%
\section{G2HDM Set Up \label{section:model}}
%%#######################################################%%

In this Section, we review the G2HDM~(cf. Ref~\cite{Huang:2015wts}) with
the particle content summarized in Table~\ref{tab:quantumnos}.

For the scalar sector, we have two Higgs doublets, $H_1$ and $H_2$, 
where $H_1$ is identified as the SM Higgs doublet and $H_2$~(with the same hypercharge $Y=1/2$ as $H_1$)
is the extra $SU(2)_L$ doublet. 
$H_1$ and $H_2$ transform as a doublet $H=(H_1\; H_2)^T$ 
under the additional gauge group $SU(2)_H\times U(1)_X$ with $U(1)_X$ charge $X(H)=1$. 
Besides the the doublet $H$, we also introduce $SU(2)_H$ triplet and doublet, $\De_H$ and $\Phi_H$, 
which are {\it singlets} under $SU(2)_L$. The Higgs potential invariant under both
$SU(2)_L\times U(1)_Y$ and $SU(2)_H \times U(1)_X$ can be written down easily as
\begin{align}
V \left( H , \Delta_H, \Phi_H \right) = V (H) + V (\Phi_H ) + V ( \De_H ) + V_{\rm mix} \left( H , \Delta_H, \Phi_H \right) \; , 
\label{eq:higgs_pot} 
\end{align}
with
\begin{align}
\label{VH1H2}
V (H) =& \; \mu^2_H H^\dag H + \la_H \lee H^\dag H \rii^2 \; ,  \nn \\    
=& \; \mu^2_H   \lee H^\dag_1 H_1 + H^\dag_2 H_2 \rii  +  \la_H \lee H^\dag_1 H_1 + H^\dag_2 H_2  \rii^2  \; , 
\end{align}
\begin{align}
\label{VPhi}
V ( \Phi_H ) =& \;  \mu^2_{\Phi}   \Phi_H^\dag \Phi_H  + \la_\Phi \lee \Phi_H^\dag \Phi_H  \rii^2  \; , \nn \\
 =& \;  \mu^2_{\Phi} \lee \Phi^*_1\Phi_1 + \Phi^*_2\Phi_2 \rii 
 +  \la_\Phi \lee \Phi^*_1\Phi_1 + \Phi^*_2\Phi_2 \rii^2 \; , \\
 \label{VDelta}
V ( \De_H ) =& \; - \mu^2_{\De} {\rm Tr} \lee \De_H^\dag \De_H  \rii  \;  + \la_\De \lee {\rm Tr} \lee \De_H^\dag \De_H  \rii \rii^2 \; , \nn \\
= & \; - \mu^2_{\De} \lee \frac{1}{2} \De^2_3 + \De_p \De_m  \rii +  \la_{\De} \lee \frac{1}{2} \De^2_3 + \De_p \De_m  \rii^2 \; , 
\end{align}
%%%%%%%%%%%%%%%%%%%%%%
and finally the mixed term
\begin{align}
\label{VMix}
V_{\rm{mix}} \left( H , \Delta_H, \Phi_H \right) = 
& \; + M_{H\De}  \lee H^\dag \De_H H \rii -  M_{\Phi\De}  \lee \Phi_H^\dag \De_H \Phi_H \rii  \nn \\
& \;  +  \la_{H\De} \lee H^\dag H  \rii    {\rm Tr} \lee \De_H^\dag \De_H  \rii  
 + \la_{H\Phi} \lee H^\dag H  \rii  \lee \Phi_H^\dag \Phi_H \rii  \nn\\
& \; + \la_{\Phi\De} \lee \Phi_H^\dag \Phi_H \rii {\rm Tr} \lee \De_H^\dag \De_H \rii  \; , \nn \\
= & \; + M_{H\De} \lee \frac{1}{\sqrt{2}}H^\dag_1 H_2 \De_p  
+  \frac{1}{2} H^\dag_1 H_1\De_3 + \frac{1}{\sqrt{2}}  H^\dag_2 H_1 \De_m  
- \frac{1}{2} H^\dag_2 H_2 \De_3   \rii   \nn \\
& \; - M_{\Phi\De} \lee  \frac{1}{\sqrt{2}} \Phi^*_1 \Phi_2 \De_p  
+  \frac{1}{2} \Phi^*_1 \Phi_1\De_3 + \frac{1}{\sqrt{2}} \Phi^*_2 \Phi_1 \De_m  
- \frac{1}{2} \Phi^*_2 \Phi_2 \De_3   \rii  \nn \\
& \; + \la_{H\De} \lee H^\dag_1 H_1 + H^\dag_2 H_2 \rii   \lee \frac{1}{2} \De^2_3 + \De_p \De_m  \rii \nn\\
& \; +  \la_{H\Phi} \lee H^\dag_1 H_1 + H^\dag_2 H_2 \rii  \lee \Phi^*_1\Phi_1 + \Phi^*_2\Phi_2 \rii \nn\\
& \; + \la_{\Phi\De}  
  \lee  \Phi^*_1\Phi_1 + \Phi^*_2\Phi_2 \rii  \lee \frac{1}{2} \De^2_3 + \De_p \De_m  \rii \; , 
\end{align}
%%%%%%%%%%
where
 \begin{align}
\De_H=
  \begin{pmatrix}
    \De_3/2   &  \De_p / \sqrt{2}  \\
    \De_m / \sqrt{2} & - \De_3/2   \\
  \end{pmatrix} \; {\rm with}
  \;\; \Delta_m = \left( \Delta_p \right)^* \; {\rm and} \; \left( \Delta_3 \right)^* = \Delta_3 \;    ,
 \end{align}
 and $\Phi_H=\left( \Phi_1 \;\; \Phi_2 \right)^T$.

Note that the quadratic terms of $H_1$ and $H_2$ have the following coefficients
\begin{equation}
\mu^2_H \mp \frac{1}{2} M_{H\De} \cdot v_\De + \frac{1}{2} \lambda_{H \De} \cdot v_\De^2 
+  \frac{1}{2} \lambda_{H \Phi} \cdot v_\Phi^2 \; ,
\end{equation} 
respectively.
As a result even with a positive $\mu_H^2$, $H_1$ can still develop a vev $(0 \; v/\sqrt 2)^T$ breaking $SU(2)_L$ provided that the second term is dominant, while $H_2$ remains zero vev.
In other words, electroweak symmetry breaking is triggered by the $SU(2)_H$ breaking.

To facilitate electroweak symmetry breaking spontaneously, it is convenience to parametrize the scalars as
\begin{eqnarray}
H_1 = 
\begin{pmatrix}
G^+ \\ \frac{v + h}{\sqrt 2} + i G^0
\end{pmatrix}
\;\; , \;\; 
\Phi_H = 
\begin{pmatrix}
G_H^p \\ \frac{v_\Phi + \phi_2}{\sqrt 2} + i G_H^0
\end{pmatrix}
\;\; , \;\; 
\Delta_H =
\begin{pmatrix}
\frac{-v_\De + \delta_3}{2} & \frac{1}{\sqrt 2}\De_p \\ 
\frac{1}{\sqrt 2}\De_m & \frac{v_\De - \delta_3}{2}
\end{pmatrix}
\end{eqnarray}
and $H_2=(H_2^+ \; H_2^0)^T$. Here $v$, $v_\Phi$ and $v_\De$ are vevs to be decided
by minimizing the potential.
$\Psi_G \equiv \{ G^+, G^3, G^p_H,  G^0_H\}$ are Goldstone bosons, to be
eaten by the longitudinal components of $W^+$, $W^3$, $W^p$, $W^{\prime 3}$ respectively, 
while $\Psi \equiv \{ h,H_2,\Phi_1,\phi_2, \de_3, \De_p \}$ are the physical fields.

Nonzero vevs $v$, $v_\Phi$ and $v_\De$ will induce the mixing among the scalars, leading to two mass matrices.
In this work, the relevant mass matrix in the basis of $\{h, \delta_3, \phi_2\}$ is given by
\begin{align}
{\mathcal M}_0^2 =
\begin{pmatrix}
2 \lambda_H v^2 & \frac{v}{2} \left( M_{H\De} - 2 \lambda_{H \De} v_\De \right) & \lambda_{H \Phi} v v_\Phi \\
\frac{v}{2} \left( M_{H\De} - 2 \lambda_{H \De} v_\De \right) & 
\frac{1}{4 v_\De} \left( 8 \lambda_\De v_\De^3 + M_{H\Delta} v^2 + M_{\Phi \De} v_\Phi^2 \right)   &  \frac{ v_\Phi}{2} \left( M_{\Phi\De} - 2 \lambda_{\Phi \De} v_\De \right) \\
\lambda_{H \Phi} v v_\Phi & \frac{ v_\Phi}{2} \left( M_{\Phi\De} - 2 \lambda_{\Phi \De} v_\De \right) & 2 \lambda_\Phi v_\Phi^2
\end{pmatrix} \; .
\label{eq:scalarbosonmassmatrix}
\end{align}

To simplify the diphoton excess analysis below, we focus on the simplest but representative scenario where
all off-diagonal terms vanish by choosing
\begin{align}
\la_{H\Phi}=0 \; , \;  M_{H\De}= 2 \la_{H\De} v_\De \; , \; M_{\Phi\De}= 2 \la_{\Phi\De} v_\De \, ,
\end{align}
and the scalar masses become
\begin{align}
m^2_h= 2 \la_H v^2  \; , \;  m^2_{\de_3}=  \frac{1}{2} \lee 4 \la_\De v^2_\De + \la_{H\De} v^2 + \la_{\Phi\De} v^2_\Phi \rii
\; , \; m^2_{\phi_2}= 2 \la_{\Phi} v^2_\Phi \, ,
\label{eq:scalar_mas}
\end{align}
where the value of $\la_H$ is exactly the same as in the SM.
In this scenario, there is no mixing among $h$, $\delta_3$, $\phi_2$\footnote{Therefore, subtleties from the scalar mixing, for example, the impact on electroweak vacuum stability~\cite{Basso:2013nza} will not be discussed here.} 
and the scalar $\phi_2$ is responsible for the diphoton excess as we shall see below. 

Next, the fermion sector together with new Yukawa couplings will be discussed.
By virtue of the additional gauge group $SU(2)_H$, new heavy fermions have to be included but there are various ways 
to implement the idea. We, however, stick to the simplest realization: the heavy fermions together with the SM right-handed fermions form $SU(2)_H$ doublets,
while the SM left-handed doublets are singlets under $SU(2)_H$.
We begin with the quark sector.
In the simplest realization, one can make the quark $SU(2)_L$ doublet, $Q_L$, an $SU(2)_H$ singlet and 
incorporate extra $SU(2)_L$ singlets $u^H_R$ and $d^H_R$ which together with the SM right-handed quarks $u_R$ and $d_R$, respectively, to form
$SU(2)_H$ doublets: $U_R^T = (u_R \;\, u^H_R)_{2/3}$ and $D_R^T = (d^H_R \;\, d_R)_{-1/3}$,
where the subscript denotes hypercharge. As a consequence, we have Yukawa couplings
\begin{align}
\mathcal{L}_{\rm Yuk} \supset & \; y_d \bar{Q}_L \lee D_R \cdot H \rii 
+ y_u \bar{Q}_L \lee U_R \cdot \stackrel{\thickapprox}{H}  \rii+ {\rm H.c.} , \nn\\
=& \; y_d \bar{Q}_L \lee d^H_R H_2 - d_R H_1  \rii - y_u \bar{Q}_L \lee u_R \tilde{H}_1 + u^H_R \tilde{H}_2  \rii   + {\rm H.c.},
\label{eq:Yuk_Q}
\end{align}
where ``$\cdot$'' refers to $SU(2)_H$ multiplication\footnote{
For 2-dimensional $SU(2)_H$ spinors $A$ and $B$, $A \cdot B = \epsilon_{ij}A^iB^j$. }
and
$\stackrel{\thickapprox}{H} \equiv ( \tilde{H}_2 \;  - \tilde{H}_1 )^T$ with 
$\tilde H_{1,2} = i \tau_2 H_{1,2}^*$ transforms as $2$ under $SU(2)_H$.
After the electroweak symmetry breaking $\lan H_1\ran \not= 0$, $u$ and $d$ obtain their masses 
but $u^H$ and $d^H$ remain massless since $H_2$ does not get a vev. 

To provide masses to the additional species, we make use of the $SU(2)_H$ scalar 
doublet $\Phi_H = (\Phi_1 \; \Phi_2)^T$, 
which is neutral under $SU(2)_L$,
and left-handed $SU(2)_{L,H}$ singlets $\chi_u$ and $\chi_d$ as
\begin{align}
\mathcal{L}_{\rm Yuk} \supset & \;  -  y^\prime_d \overline{\chi}_d \lee D_R \cdot \Phi_H \rii 
+ y'_u \overline{\chi}_u \lee U_R \cdot \tilde{\Phi}_H \rii  + {\rm H.c.} , \nn\\
=& \;  -  y'_d \overline{\chi}_d \lee d^H_R \Phi_2 - d_R \Phi_1  \rii 
- y'_u \overline{\chi}_u \lee u_R \Phi^*_1 + u^H_R \Phi^*_2  \rii   + {\rm H.c.}, 
\label{eq:Yuk_Q1} 
\end{align}
in which $\Phi_H$ has $Y=0$, $Y(\chi_u)=Y(U_R)=2/3$ and $Y(\chi_d)=Y(D_R)=-1/3$ with
$\tilde{\Phi}_H=( \Phi^*_2 \; - \Phi^*_1 )^T$. With $\lan \Phi_2 \ran= v_{\Phi}/\sqrt{2}$, $u^H~(\chi_u)$ and $d^H~(\chi_d)$ obtain masses
$y'_u v_{\Phi}/\sqrt{2}$ and $y'_d v_{\Phi}/\sqrt{2}$, respectively. 
Notice that both $v_\De$ and $v_{\Phi}$ contribute to the $SU(2)_H$ gauge boson masses.
 
The lepton sector mimics the quark sector as
\begin{align}
\mathcal{L}_{\rm Yuk} \supset & \;  y_e \bar{L}_L \lee E_R \cdot H \rii 
+ y_\nu {\bar L}_L \lee N_R \cdot \tilde H \rii 
 -  y'_e \overline{\chi}_e \lee E_R \cdot \Phi_H \rii   
+ y'_\nu \overline{\chi}_\nu \lee N_R \cdot \tilde \Phi_H \rii   
+ {\rm H.c.} , \nn\\
=& \; y_e \bar{L}_L \lee e^H_R H_2 - e_R H_1 \rii 
- y_\nu \bar{L}_L \lee \nu_R \tilde{H_1} + \nu^H_R \tilde{H_2} \rii \nn\\
& \; -  y'_e \overline{\chi}_e \lee e^H_R \Phi_2 - e_R \Phi_1  \rii   
- y'_\nu \overline{\chi}_\nu \lee \nu_R \Phi^*_1 + \nu^H_R \Phi^*_2 \rii
+ {\rm H.c.},
\label{eq:Yuk_L}
\end{align}
in which $E_R^T = (e^H_R  \; e_R)_{-1}$ and $N_R^T = (\nu_R  \;  \nu^H_R)_{0}$ where $\nu_R$ and $\nu^{H}_R$
correspond to the right-handed neutrino and the $SU(2)_H$ partner of it respectively, 
while $\chi_e$ and $\chi_\nu$ are $SU(2)_{L,H}$ singlets with $Y(\chi_e)=-1$ and $Y(\chi_\nu)=0$.
Similarly all SM leptons and their heavy counterparts will obtain masses from $\langle H_1 \rangle$ and
$\langle \Phi_2 \rangle$.

As mentioned above, because $\phi_2$~(a member of $\Phi_H$) couples to the new heavy fermions, 
it can be radiatively produced via loops of the new colored particles and radiatively decays into the diphoton final state 
via loops of the new charged particles to accommodate the observed bump.
On the other hand, although $\phi_2$ is a singlet under the SM gauge group, it does couple to SM fermions and gauge bosons
at tree level via the $h-\phi_2$ mixing. That is the reason why we work in the zero mixing limit to evade direct search bounds from, for instance, dijet or dilepton channels. Note that there are no excesses in the $ZZ$, dijet or dilepton 
channels near the invariant mass of 750 GeV.

\begin{table}[htdp!]
\begin{tabular}{|c|c|c|c|c|c|}
\hline
Matter Fields & $SU(3)_C$ & $SU(2)_L$ & $SU(2)_H$ & $U(1)_Y$ & $U(1)_X$ \\
\hline \hline
$Q_L=\left( u_L \;\; d_L \right)^T$ & 3 & 2 & 1 & 1/6 & 0\\
$U_R=\left( u_R \;\; u^H_R \right)^T$ & 3 & 1 & 2 & 2/3 & $1$ \\
$D_R=\left( d^H_R \;\; d_R \right)^T$ & 3 & 1 & 2 & $-1/3$ & $-1$ \\
\hline
$L_L=\left( \nu_L \;\; e_L \right)^T$ & 1 & 2 & 1 & $-1/2$ & 0 \\
$N_R=\left( \nu_R \;\; \nu^H_R \right)^T$ & 1 & 1 & 2 & 0 & $1$ \\
$E_R=\left( e^H_R \;\; e_R \right)^T$ & 1 & 1 & 2 &  $-1$  &  $-1$ \\
\hline
$\chi_u$ & 3 & 1 & 1 & 2/3 & 0 \\
$\chi_d$ & 3 & 1 & 1 & $-1/3$ & 0 \\
$\chi_\nu$ & 1 & 1 & 1 & 0 & 0 \\
$\chi_e$ & 1 & 1 & 1 & $-1$ & 0 \\
\hline\hline
$H=\left( H_1 \;\; H_2 \right)^T$ & 1 & 2 & 2 & 1/2 & $1$ \\
$\Delta_H=\left( \begin{array}{cc} \Delta_3/2 & \Delta_p/\sqrt{2}  \\ \Delta_m/\sqrt{2} & - \Delta_3/2 \end{array} \right)$ & 1 & 1 & 3 & 0 & 0 \\
$\Phi_H=\left( \Phi_1 \;\; \Phi_2 \right)^T$ & 1 & 1 & 2 & 0 & $1$ \\
\hline
\end{tabular}
\caption{Matter field contents and their quantum number assignments in G2HDM. 
}
\label{tab:quantumnos}
\end{table}

%%#######################################################%% 

\section{Diphoton Anomaly} \label{section:diph}

%%#######################################################%%  
Equipped with the basics of G2HDM, we are now in a position to calculate the diphoton cross section via $\phi_2$ exchange.  
The cross section at the $\phi_2$-resonance can be well approximated by~\cite{Leike:1998wr} 
\begin{align}
\sig\lee gg \to \phi_2 \to \ga\ga \rii = \frac{ \pi^2}{ 8 s \, m_{\phi_2} \, \Ga_{\phi_2} } f_{gg} \lee \frac{m_{\phi_2}}{\sqrt{s}} \rii
 \Ga\lee \phi_2 \to gg \rii \Ga\lee \phi_2 \to \ga\ga \rii ,
\end{align}
with the center of mass energy $\sqrt{s}=13$ TeV and the integral of the parton~(gluon in this case) distribution function product 
\begin{align}
 f_{gg} = \int^1_{m^2_{\phi_2}/s} \frac{d x}{x} g\lee x, \mu^2 \rii g\lee \frac{ m^2_{\phi_2} } {s x}, \mu^2 \rii = 2141.7,
\end{align}
evaluated at the scale $\mu=m_{\phi_2}$, using MSTW2008NNLO~\cite{Martin:2009iq} and
the value is consistent with Ref.~\cite{Franceschini:2015kwy}.
The partial decay width of $\phi_2$ into a heavy fermion and antifermion  in the presence of a Yukawa term,
$y^\prime_f  \phi_2 \bar{f} f /\sqrt{2}$, that also gives a mass $m_{f}$ to the heavy fermion because of $\lan \phi_2\ran = v_\Phi$, reads
\begin{align}
\Ga \lee \phi_2 \to f \bar{f}\rii = N_c \frac{y^{\prime2}_f \, m_{\phi_2} }{16 \pi} \lee 1 - 4 \frac{ m^2_{f}}{m^2_{\phi_2}} \rii^{3/2} \, ,
\label{eq:phi2de}
\end{align}
where $N_c=3$ for heavy colored particles while $N_c=1$ for heavy leptons. 

The partial decay width of $\phi_2$ into diphoton mediated by heavy fermions is~\cite{Gunion:1989we,Djouadi:2005gi,Djouadi:2005gj}
\begin{eqnarray}
\Gamma \, (\phi_2\to \gamma\gamma) = \frac{ \alpha^{2}\,  m^3_{\phi_2} }
{256 \, v^2_{\Phi}  \,\pi^{3}} 
\left| \sum_{f}  N_{c} Q_f^2 A_{1/2}^{H}(\tau_f) 
\right|^2 , \nn \\
\label{eq:hgagaH2}
\end{eqnarray}
where $\tau_f= m^2_{\phi_2}/4 m^2_f $ with
\begin{eqnarray}
A_{1/2}^{H}(\tau) & = & 2 [\tau +(\tau -1)f(\tau)]\, \tau^{-2}  \, ,
\label{eq:A0+Af+Aw}
\end{eqnarray}
and the function $f(\tau)$ is defined as
\begin{eqnarray}
f(\tau)=\left\{
\begin{array}{ll}  \displaystyle
\arcsin^2\sqrt{\tau} & {\rm , \;\; for} \; \tau\leq 1 \; ; \\
\displaystyle -\frac{1}{4}\left[ \log\frac{1+\sqrt{1-\tau^{-1}}}
{1-\sqrt{1-\tau^{-1}}}-i\pi \right]^2 \hspace{0.5cm} & {\rm , \; \; for} \; \tau>1 \; .
\end{array} \right.
\label{eq:ftau}
\end{eqnarray}

On the other hand, the partial decay width of $\phi_2$ into 2 gluons mediated by colored heavy fermions is~\cite{Gunion:1989we,Djouadi:2005gi,Djouadi:2005gj}
\begin{eqnarray}
\Gamma \, (\phi_2\to g g) = \frac{ \alpha^{2}_s\,  m^3_{\phi_2} }
{72 \, v^2_{\Phi}  \,\pi^{3}} 
\left| \sum_{f}  \frac{3}{4} A_{1/2}^{H}(\tau_f) 
\right|^2 \; .
\label{eq:hgagaH2}
\end{eqnarray}

In our model, there are 6 heavy colored Dirac fermions, including 3 generations of up-type and down-type heavy quarks~(with electric charge of 
$2/3$ and $1/3$, respectively) which contribute in $\Gamma \, (\phi_2\to g g)$ while for $\Gamma \, (\phi_2\to \ga\ga)$ there are additional 3 heavy charged
leptons with one unit of electric charge in addition to the heavy quarks.
From the CMS run I and CMS+ATLAS run II diphoton data combined, the best fit value for the diphoton cross section is $6.2\pm 1.0$ femtobarn~\cite{Ellis:2015oso}.
It implies in units of GeV$^{-2}$
\begin{align}
\sig\lee gg \to \phi_2 \to \ga\ga \rii  = \frac{ f_{gg} \lee \frac{m_{\phi_2}}{\sqrt{s}} \rii \pi^2}{ 8 s } \frac{m_{\phi_2}}{\Ga_{\phi_2}}
 \frac{\Ga\lee \phi_2 \to gg \rii}{m_{\phi_2}}
 \frac{\Ga\lee \phi_2 \to \ga\ga \rii}{m_{\phi_2}} \simeq 1.60 \times 10^{-11} ,
\end{align}
{\it i.e},
\begin{align}
1.65 \times 10^{-8} \simeq  \frac{\Ga\lee \phi_2 \to gg \rii}{m_{\phi_2}}
 \frac{\Ga\lee \phi_2 \to \ga\ga \rii}{m_{\phi_2}} ,
\end{align}   
with $\sqrt{s}=13$ TeV and $\Ga_{\phi_2}/m_{\phi_2} \simeq 0.06$~\cite{ATLASga}.

\begin{figure}[htbp!]
\centering
\includegraphics[clip,width=0.60\linewidth]{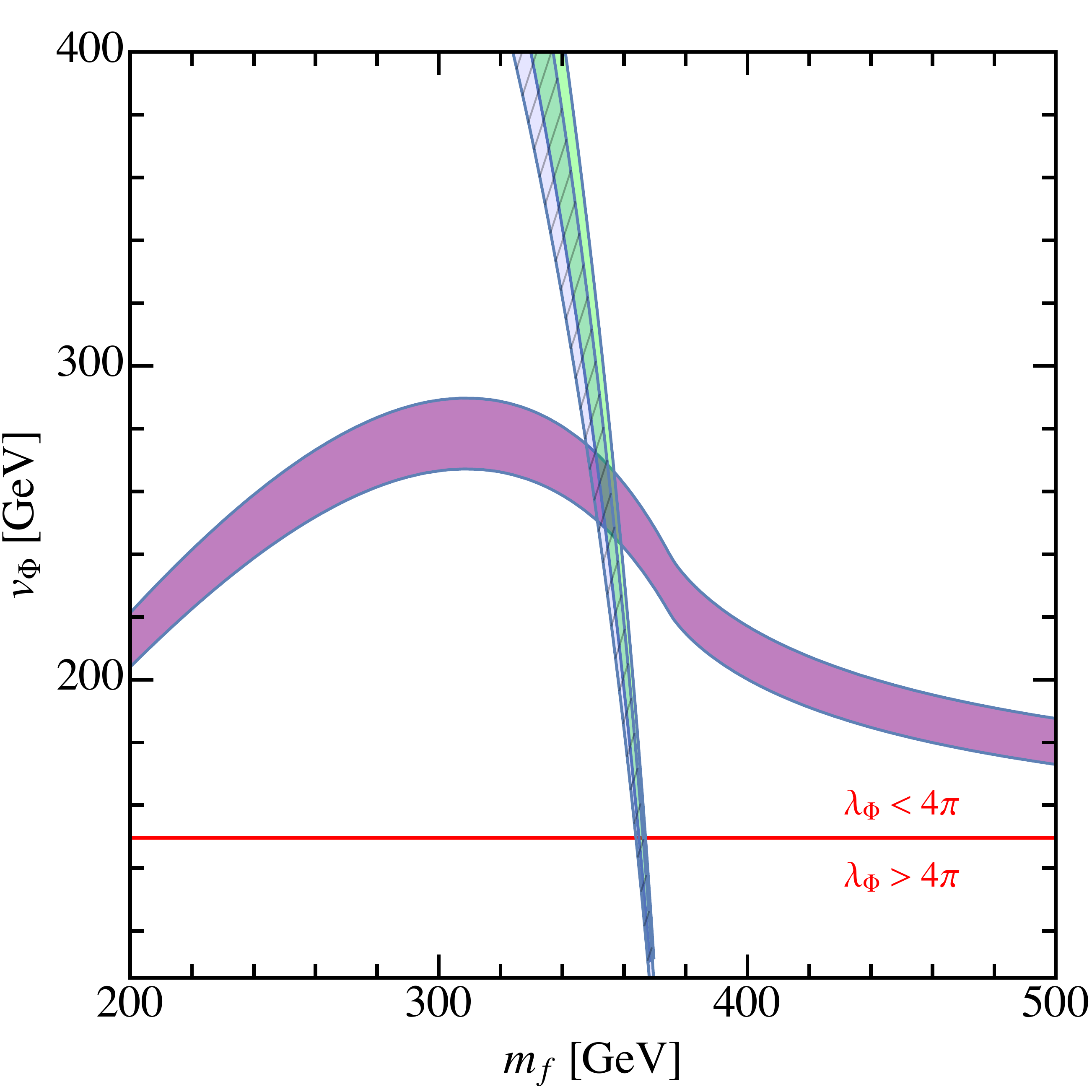}
\caption{ The purple area on the $m_f - v_\Phi$ plane is the $1\sigma$ region which reproduces the $\ga\ga$ bump at the LHC. The green shaded region denotes $0.05<\Ga_{\phi_2}/m_{\phi_2} <0.07$, including all neutral and charged heavy fermions, while the blue shaded region takes into account the heavy charged particles only. 
The red solid line marks the perturbativity limit because $m^2_{\phi_2} = 2 \la_{\Phi} v^2_{\Phi}$.
In order to reproduce the diphoton bump with the proper width, one will
need fermion masses to be around 360 GeV and the vev $v_\Phi$ at 250 GeV, implying $\mathcal{O}(1)$ Yukawa couplings.
}
\label{fig:mf_vPhi}  
\end{figure}

In the Fig.~\ref{fig:mf_vPhi}, we color the $1\sigma$ region in purple 
on the $m_f - v_\Phi$ plane to accommodate the $\ga\ga$ anomaly 
where all heavy fermions involved are assumed to have the same mass $m_f$ for simplicity.
The green shaded region corresponds to the total decay width of $\phi_2$, obtained from Eq.~\eqref{eq:phi2de} by including all neutral and charged heavy fermions ($u^H,d^H,e^H,\nu^H$),
at the range of $0.05<\Ga_{\phi_2}/m_{\phi_2} <0.07$ that is consistent with the observed resonance width~\cite{ATLASga}.
By contrast, the blue shaded region denotes the total decay width of $\phi_2$ with 
$0.05<\Ga_{\phi_2}/m_{\phi_2} <0.07$,
including heavy charged particles only ($u^H,d^H,e^H)$. 
The red solid line corresponds to the perturbativity limit since $m^2_{\phi_2} = 2 \la_{\Phi} v^2_{\Phi}$ in the limit of zero mixing among $h$, $\phi_2$ and $\de_3$.
In order to have the diphoton excess, one can see that 
the new fermion masses have to be around 360 GeV with the vev $v_\Phi$ at 250 GeV.
However,  we can also relax our assumption to allow for non-degenerate heavy fermion masses.
In this case, one can still achieve the diphoton excess and the desired total decay width of $\phi_2$,
while the heavy charged fermion masses are not longer constrained to be around 360 GeV.

We conclude this Section by commenting on impacts of having $v_\Phi$ around 250 GeV.
As discussed in Ref.~\cite{Huang:2015wts}, $v_\Phi$ is restrained to be of order TeV to avoid various constraints.
Small $v_\Phi$ will induce a large mixing between the SM $Z$ and $SU(2)_H$ $Z^\prime$,
which can be avoided
if the $SU(2)_H$ gauge coupling $g_H$ is small. To be more clear, the mixing angle, in the limit of $g_H \ll g$, reads
\begin{align}
\sin\th_{ZZ^\prime} \simeq - \frac{g_H}{\sqrt{ g^2 + g^{\prime} } },
\label{eq:z-zmixing}
\end{align}
where $g$ and $g^\prime$ are the SM $SU(2)_L$ and $U(1)_Y$ gauge coupling constants, respectively.
One can in principle make $g_H$ small to have a very small mixing, resulting in very light $SU(2)_H$ gauge bosons.
On the other hand, the DM matter candidate in this case could be the new neutral lepton ($\nu^H_R$ or $\chi_\nu$),
the $SU(2)_H$ $W^\prime$ or the neutral Higgs $H^0_2$, depending on the parameter space.
The DM stability is protected by the $SU(2)_H$ gauge symmetry and the Lorentz invariance as demonstrated in Ref.~\cite{Huang:2015wts}.

%%#######################################################%% 
\section{Implications of a few hundred GeV heavy fermions} \label{section:Hfermion}
%%#######################################################%%  

In this Section, we briefly comment on some of consequences of $SU(2)_H$ heavy fermions with masses of order 
360 GeV, required to realize the diphoton excess. A detailed study is, however, beyond the
scope of this paper and deserves a separate work.

\subsection{Muon and Electron magnetic dipole moment $g-2$}

At one-loop level, the charged leptons~(electron and muon) anomalous magnetic moment $(g_\ell - 2)$ receive three additional radiative contributions\footnote{To simplify the analysis, we treat $U(1)_X$ as a global symmetry by setting $g_X=0$.}
involving loops of $W^\prime$ with $\ell^H$, $H_2$ with $\ell^H$ and $Z^\prime$ with $\ell^H$, out of which the $H_2$ contribution can be neglected because it is highly suppressed by the corresponding small SM
electron and muon Yukawa couplings and $H_2$ are assumed to be heavy.
Taking into account the fact $W^\prime$ and $Z^\prime$ only couple to the right-handed SM fermions,
the gauge boson contributions to the anomaly $a_\ell \equiv (g_\ell -2)/2$ are~\cite{Murakami:2001cs,Pospelov:2008zw}
\begin{align}
a_l^{W^\prime}  &= \frac{g^2_H}{ 32 \pi^2 } \int^1_0 dx \,
\frac{ \lee 1 - x \rii }{  r^2_{W^\prime}  \lee r^2_{H} \lee 1 -x \rii  + \lee r^2_{W^\prime} - \lee 1 -x \rii  \rii  x \rii }  \nonumber\\
& \;\; \; \times \lee r_{H} \lee 1 -x \rii^3 + 4 r_{H} r^2_{W^\prime} x 
+  \lee 1 - x\rii^2 x - \lee r^2_{H} \lee1-x\rii^2 + 2 r^2_{W^\prime} x \lee 1 + x \rii\rii \rii \nonumber\\
  &\simeq \frac{g^2_H}{ 48 \pi^2 } \left\{
\begin{array}{l l} 			\frac{ 12 r_{H} r^2_{W^\prime} - 9  r^2_{W^\prime} - 2 r^2_{H}  }{ 4 r^2_{H} r^2_{W^\prime} }		
						  &     ~~{\rm for} ~~ \mlh \gg m_{W^\prime}  > m_{\ell} \;\; ,\\
						       \frac{ 3 r_{H} - 2 }{  r^2_{W^\prime} } & ~~{\rm for} ~~  m_{W^\prime} \gg \mlh > m_{\ell} \;\; ,
										\end{array}
\right.
\label{eq:g-2Wp}
\end{align}
and 
\begin{align}
a_l^{Z^\prime}  &= \frac{g^2_H}{ 32 \pi^2 } \int^1_0 dx \, \frac{ x \lee 1 - x \rii^2 }{ \lee 1 - x\rii^2 + r^2_{Z^\prime} x }
 \nonumber \\
&\simeq \frac{g^2_H}{64 \pi^2 } \left\{
\begin{array}{l l} 			1  &     ~~{\rm for} ~~ m_{\ell} \gg m_{Z^\prime} \;\; ,\\
						      \frac{ 2 }{3 r^2_{Z^\prime}}  & ~~{\rm for} ~~ m_{Z^\prime} \gg  m_{\ell} \;\; ,
										\end{array}
\right.
\label{eq:g-2Zp}
\end{align}
where $r_{H} \equiv \mlh/m_\ell$ and $r_{(W^\prime ,Z^\prime)} \equiv  m_{(W^\prime,Z^\prime)}/m_{\ell}$.

\begin{figure}[htp!]
\centering
\includegraphics[clip,width=0.46\linewidth]{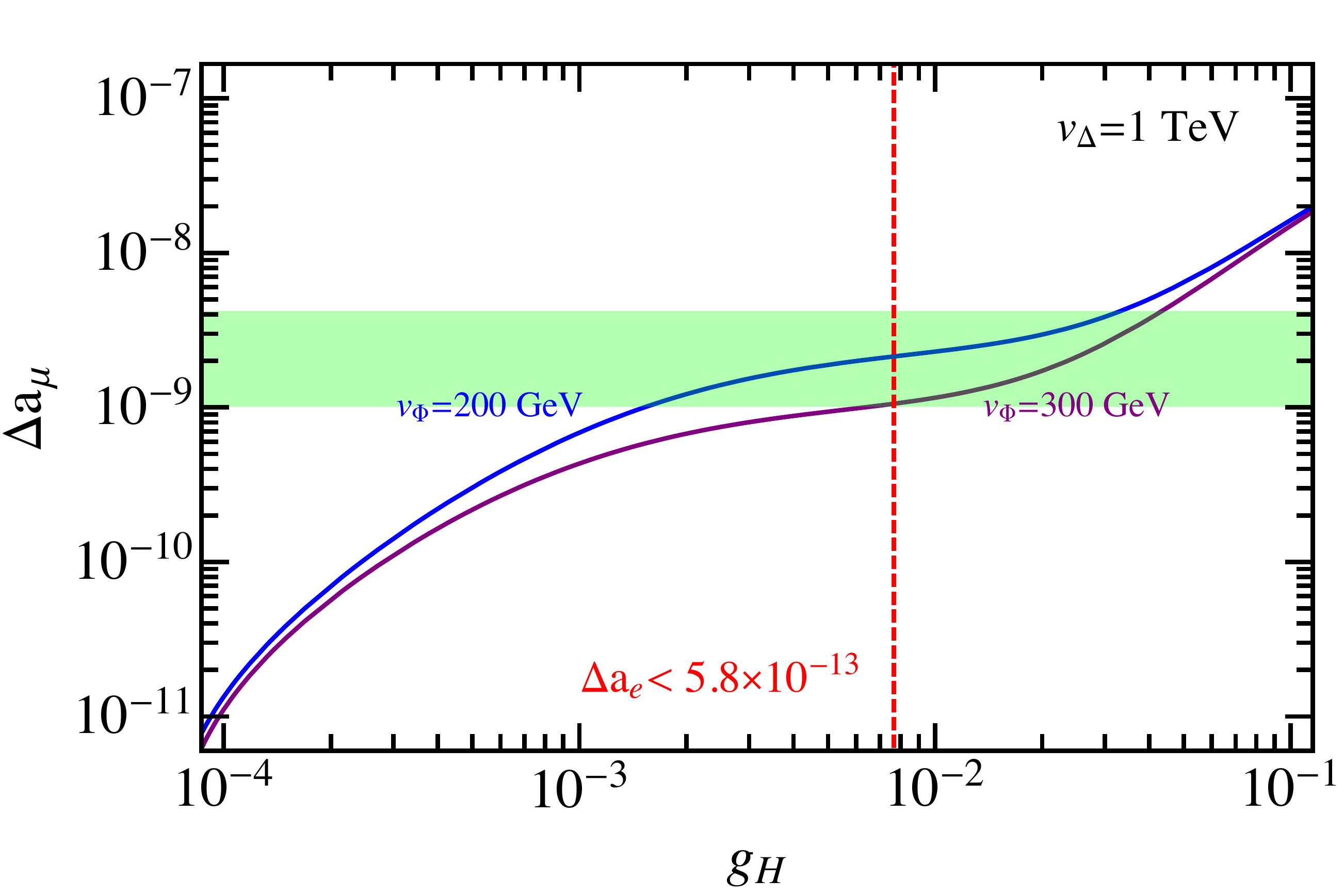}
\includegraphics[clip,width=0.46\linewidth]{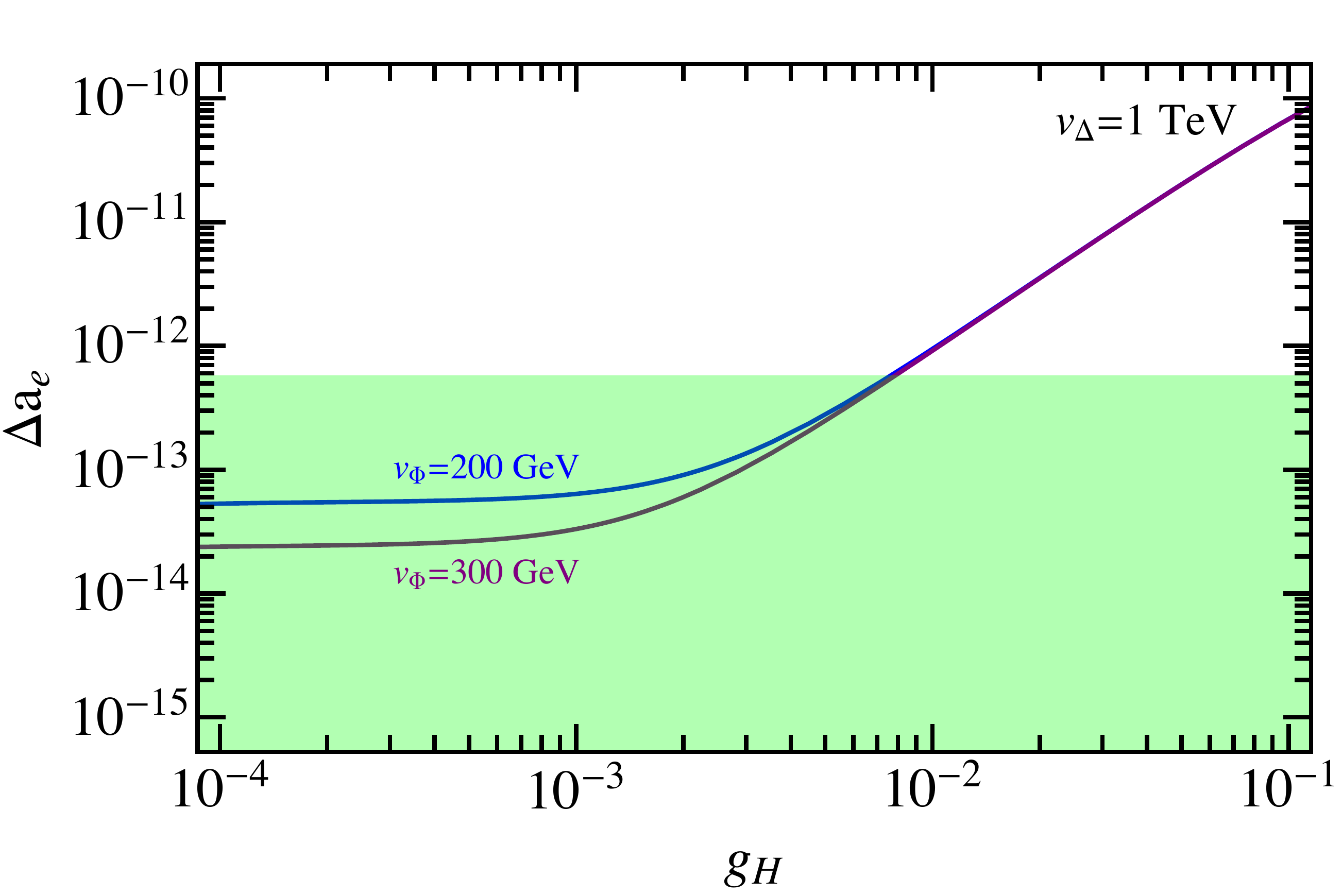}
\caption{ Muon and electron  $\Delta a_\ell \equiv ( \Delta g_\ell -2)/2$, where $\Delta g_\ell$ refers to extra contributions
to the magnetic dipole moment from G2HDM. }
\label{fig:g-2}  
\end{figure}

In addition, the $Z-Z^\prime$ mixing with the angle given in Eq.~\eqref{eq:z-zmixing} also induces an extra  contribution to $a_l$, obtained by multiplying Eq.~\eqref{eq:g-2Zp} by $\lee \sin {\th_{Z Z'}} \rii^2$ and replacing $g_H$ by $g/(\cos\th_w)$, where $\th_w$
is the Weinberg angle. In contrast, due to the quantum number assignment, $W^\prime$ is electrically neutral
and will not mix with the SM $W$ boson, unlike $Z^\prime$. 
Thus, Eq.~\eqref{eq:g-2Wp} is the total contribution from $W^\prime$.
Moreover, the $W^\prime$ and $Z^\prime$ boson masses are
\begin{align}
m^2_{W^{\prime}}  &= \frac{1}{4} g^2_H \lee v^2 + v^2_\Phi + 4 v^2_\De \rii  \; , \nonumber \\ 
m^2_{Z^\prime} &\simeq  \frac{1}{4} g^2_H v^2_\Phi \; ,  \; \; \text{ ( in the limit of $g_H \ll g, g^\prime$ ) } \; .
\end{align}
We present our results in Fig.~\ref{fig:g-2} where all of $Z^\prime$, $W^\prime$ and $Z-Z^\prime$ mixing contributions are included.
In the left-panel, with $v_\Delta$ set to 1 TeV and $\mlh$ to be 360 GeV, the green band on the $g_H - \Delta a_\mu$ plane corresponds to the $2\sigma$ region of the difference between the experimental value and the SM prediction~\cite{Bennett:2006fi,Davier:2010nc,Hagiwara:2011af}, $ 10.1 \times 10^{-10} <a^{\text{exp}}_{\mu} - a^{\text{SM}}_{\mu}<  42.1\times 10^{-10}$, the blue~(purple) line refers to $v_{\Phi}=200~(300)$ GeV. To explain the muon anomaly $\Delta a_\mu$, small values of $v_\Phi$ are preferred. 
The red dashed line is the limit extracted from the electron anomaly $\Delta a_e$ 
as shown in the right panel, where the green band represents
$-2.7 \times 10^{-12} <a^{\text{exp}}_{e} - a^{\text{SM}}_{e}<  5.8 \times 10^{-13}$~\cite{2008PhRvL.100l0801H,Hanneke:2010au,Aoyama:2012wj,Endo:2012hp}.
For $g_H \lesssim 10^{-3}$, the electron anomaly $\Delta a_e$ scales as $g^2_H m^2_e/m^2_{(W^\prime, Z^\prime)}$, which is simply
$ m^2_e/v^2_{(\Delta,\Phi)}$ since $m^2_{(W^\prime, Z^\prime)} \sim g_H^2 v^2_{(\Delta,\Phi)}$. This implies independence of $\Delta a_e$ on $g_H$. 
However, for $g_H \gtrsim 10^{-2}$ it is proportional to $g^2_H$,  since for
$\mlh \sim m_{W^\prime} \gg m_\ell $, $a^{W^\prime}_e \sim g_H \frac{m_\ell}{\mlh}$ from Eq.~\eqref{eq:g-2Wp}.

\subsection{Collider Searches}

In previous subsection, we showed  that in order to accommodate the diphoton excess without contradicting the electron and muon $g-2$ measurement,
the $SU(2)_H$ gauge coupling $g_H$ is confined to be less than $10^{-2}$. Thus, at the LHC the heavy fermions will be mainly produced
via the 750 GeV $\phi_2$ decay due to large Yukawa couplings of $\mathcal{O}(1)$ instead of being generated through $W^\prime$- and
$Z^\prime$-exchange processes. By virtue of the $SU(2)_H$ gauge symmetry, the decay of these heavy fermions must be accompanied by the DM particle in the final state as well.

For illustration, we use $\tau^H$ as an example. It has three different decay channels, corresponding to three possible DM candidates
$\nu^H$, $H_2^0$ and $W^\prime$ in G2HDM, respectively:
\begin{align}
\tau^H & \to W^{\prime p} \, \tau_R \to \nu^H \, \overline{\nu_R} \, \tau_R \, , \nonumber \\ 
\tau^H & \to H_2^0 \, \tau_L \, ,  \nonumber \\ 
\tau^H & \to W^{\prime p} \, \tau_R \, ,
\end{align}  
where in the first channel one could have multiple leptons or jets
in addition to missing energy depends on whether $\overline{\nu_R}$ decays 
into $\overline{\nu_L}$ and $H_1$ within the detector or not, while
the last two channels feature one lepton plus missing transverse energy.

The energy of SM fermion $\tau$ in the final state 
depends on the mass difference between $\tau^H$ and 
the DM. If the mass splitting is too small, this may lead
to very soft $\tau$ which fails to pass the event selection. The process $gg \to \phi_2 \to \overline{\tau^H} \tau^H \to
\text{null}~(\text{DM} + \text{soft $\tau$s}) $,  which will be largely excluded  by the DM mono-jet searches as pointed out in Ref.~\cite{Han:2016pab}.
On the other hand, if the mass splitting is large enough, the final state $\tau$ is 
visible and the situation will require delicate study, see Ref.~\cite{Han:2016pab} 
for more details.

%since for $m_{\tau^H} \sim 360$ GeV~(in light of the diphoton excess with a large width) $\tau$ will have energy of a few hundred GeV, which could potentially be overwhelmed by background, especially from the QCD background. 
%Besides, one can not reconstruct the invariant mass of $\tau^H$ from adding up the SM fermion momenta because of the existence of
%DM which manifests as missing energy.

%It is worthy of mentioning that 
%the new lepton is possibly detectable in LHC 
%by decaying into a SM fermion plus a DM particle,
%
%\begin{equation}
%f^{H}\to\chi+f^{SM}, 
%\nonumber
%\end{equation}
%where DM $\chi$ can be inert scalar ($H_{2}^{0}$) or 
%fermion ($\nu_{R}^{H}$ or $\chi_{\nu}$).  
%\st{Please help me the spin conservation of fermion $\chi$!!!}
%However, if the mass splitting between the DM and 
%new lepton is small then the SM fermion energy 
%can be very soft, failing to pass the event selection.
%\mkblue{On the other hand, if the mass splitting is too small, 
%a large missing energy can be detected by measuring 
%the initial radiation mono-jet. In Ref.~\cite{Han:2016pab}, 
%the required mass splitting between DM and new lepton for the 
%various new fermion masses are given in their Fig. 2.}

\subsection{Electroweak Precision Test  - $\Delta S$, $\Delta T$ and $\Delta U$}

Finally, we would like to comment on extra corrections from additional particles in G2HDM
to the electroweak oblique observables.
In additional to the SM particles,  G2HDM contains the new $SU(2)_L$ doublet $H_2$, the $SU(2)_H$
gauge bosons of which $Z^\prime$
mixes with the SM $Z$, and the heavy $SU(2)_H $ fermions. Other scalars $\Phi_H$ and $\Delta_H$ are singlets under $SU(2)_L$
and hence are not relevant.

The heavy fermions, as $SU(2)_L$ singlets, will not contribute to electroweak corrections described by the oblique parameters,
$\Delta S$, $\Delta T$ and $\Delta U$, as can be easily seen from the definition of the parameters~\cite{Peskin:1991sw}.
Moreover, as demonstrated above the $Z-Z^\prime$ mixing is constrained by the electron $g-2$ bound to be less than $10^{-2}$ or so, implying contributions
to the oblique parameters at the order of $10^{-4}$ or smaller. Finally as long as the mass splitting between $H_2^\pm$ and $H^0_2$
is small, corrections to $\Delta S$, $\Delta T$ and $\Delta U$ will be suppressed~\cite{Barbieri:2006dq}. All in all, this model can survive from the electroweak
precision test.

%%%%%%%%%%%%%%%%%%%%%%%%%%%%%%%%%%%
  
\section{Conclusion} \label{section:conclusion}

In this work, we address a possible solution to the diphoton anomaly observed at the LHC based on the recent G2HDM model proposed by us.
In the G2HDM, the two Higgs doublets $H_1$ and $H_2$ are embedded into a doublet under
a non-abelian gauge symmetry $SU(2)_H$ and the resulting 
$SU(2)_H$ doublet is charged under an additional abelian group $U(1)_X$.
To give  masses to additional gauge bosons, we introduce a $SU(2)_H$ scalar triplet and a 
doublet~(both are singlets under the SM
gauge group). On the other hand, extra new heavy fermions 
are needed to have Yukawa couplings comply with the $SU(2)_H$ gauge symmetry.
In other words, we have only {\it chiral} fermions, different from some of existing models where vector-like quarks and leptons are employed to explain the anomaly.  
In addition, constraints on new vector-like quarks and leptons because of mixing with SM fermions~\cite{delAguila:2008pw,delAguila:2008cj,Aguilar-Saavedra:2013qpa} do not apply here since our new fermions do not mix with the SM ones.

The new heavy fermions receive masses from the vev of the $SU(2)_H$ scalar doublet, that also provides masses to the additional gauge bosons.
A physical component $\phi_2$ inside the doublet can be produced radiatively via gluon fusion with the additional heavy colored fermions in loops
and in turn radiatively decays into two photons with the heavy charged fermions involved. 
We have shown that in the limit of the universal fermion mass,
in order to reproduce the anomaly, the vev of $\phi_2$ ranges from 180 to 300 GeV 
with the new fermion mass of few hundred GeV.
The desired total decay width of $\Ga_{\phi_2} \simeq 0.06 m_{\phi_2}$, by having $\phi_2$ decay into the new fermions, can be realized with $m_f\sim 360$ GeV and $v_\Phi \sim 250$ GeV.
The favorable region could be further extended if the additional neutral fermions are allowed to have arbitrary masses.

The existence of $SU(2)_H$ gauge bosons can also explain the anomalous muon magnetic dipole moment.
There are three radiative corrections to muon $g-2$: $W^\prime$ with $\mu^H$, $Z^\prime$ with $\mu_R$ and the correction
induced by the $Z-Z^\prime$ mixing. We have found out with $m_{\mu^H} =360$ GeV and $g_H \sim 7\times 10^{-3}$, resulting in 
GeV or sub-GeV $W^\prime$ and $Z^\prime$ depending on the vevs of $\Phi_H$ and $\Delta_H$, the muon anomaly $\Delta a_\mu$ of order $10^{-9}$ can be realized
while the corresponding contributions to electron anomaly $\Delta a_e$ is highly suppressed by the very small electron mass.

We conclude by pointing out that except for the diphoton anomaly, the LHC run-II data do not feature any significant deviation from the SM prediction.
Our model can avoid overproducing other SM model particles through the same $\phi_2$ exchange process since $\phi_2$ couples only to the extra fermions at tree level in the limit of the vanishing $h-\phi_2$ mixing.
The heavy fermions from $\phi_2$ decays, however, subsequently decay into SM particles plus the DM particles, 
that manifest as missing transverse energy.
The resulting SM particle energy spectra depend on the mass difference between the new heavy fermions and DM, and the spectra could be very soft if the mass difference is small
just like the compressed spectra in various supersymmetry models.         
Finally,  for the zero $h-\phi_2$ mixing, one can expect the $Z \ga$ and $ZZ$
signals with a similar order of magnitude as in the $\ga\ga$ anomaly 
through the same $\phi_2$ exchange process. 

\vfill 
%\newpage

\section*{Acknowledgments}
WCH would like to thank Ivan Ni\v sand\v zi\' c for very useful discussion on parton distribution functions.
This work is supported in part by the Ministry of Science and Technology (MoST) of Taiwan under
grant number 104-2112-M-001-001-MY3 (TCY), DGF Grant No. PA 803/10-1 (WCH), 
and the World Premier International Research Center Initiative (WPI), MEXT, Japan (YST).

\bibliography{SU2_diph}
\bibliographystyle{h-physrev}

\end{document}